\documentclass[12pt]{article}

\usepackage{amsfonts}

\usepackage{amssymb}
\usepackage{latexsym}
\usepackage{graphicx}
\usepackage[english]{babel}
\usepackage[font={small}]{caption}

\usepackage{amsfonts}
\usepackage{latexsym}
\usepackage{graphicx}
\usepackage[english]{babel}
\usepackage{tikz}

\begin{document}
\input epsf

\def\p{\partial}
\def\h{{1\over 2}}
\def\be{\begin{equation}}
\def\bea{\begin{eqnarray}}
\def\ee{\end{equation}}
\def\eea{\end{eqnarray}}
\def\d{\partial}
\def\la{\lambda}
\def\eps{\epsilon}
\def\bb{\bigskip}
\def\mm{\medskip}
\newcommand{\dm}{\begin{displaymath}}
\newcommand{\edm}{\end{displaymath}}
\renewcommand{\b}{\tilde{B}}
\newcommand{\gm}{\Gamma}
\newcommand{\ac}[2]{\ensuremath{\{ #1, #2 \}}}
\renewcommand{\ell}{l}
\newcommand{\z}{\ell}
\newcommand{\newsection}[1]{\section{#1} \setcounter{equation}{0}}
\def\bb{$\bullet$}
\def\Qbar{{\bar Q}_1}
\def\QPbar{{\bar Q}_p}

\def\q{\quad}

\def\bn{B_\circ}

\let\a=\alpha \let\b=\beta \let\g=\gamma \let\d=\delta \let\e=\epsilon
\let\c=\chi \let\th=\theta  \let\k=\kappa
\let\l=\lambda \let\m=\mu \let\n=\nu \let\x=\xi \let\r=\rho
\let\s=\sigma \let\t=\tau
\let\vp=\varphi \let\vep=\varepsilon
\let\w=\omega      \let\G=\Gamma \let\D=\Delta \let\Th=\Theta
                     \let\P=\Pi \let\S=\Sigma

\def\h{{1\over 2}}
\def\t{\tilde}
\def\r{\rightarrow}
\def\nn{\nonumber\\}
\let\bm=\bibitem
\def\Kt{{\tilde K}}
\def\b{\bigskip}

\let\p=\partial

\begin{flushright}
\end{flushright}
\vspace{20mm}
\begin{center}
{\LARGE What prevents gravitational collapse \\
in string theory?   \footnote{Essay awarded an honorable mention in  the Gravity Research Foundation 2016  essay competition.}}
\\
\vspace{18mm}
 Samir D. Mathur

\vskip .1 in

 Department of Physics\\The Ohio State University\\ Columbus,
OH 43210, USA\\mathur.16@osu.edu\\
\vspace{4mm}
 March 31, 2016
\end{center}
\vspace{10mm}
\thispagestyle{empty}
\begin{abstract}

It is conventionally believed that if a ball of matter of mass $M$ has a radius close to $2GM$ then it must collapse to a black hole. But string theory  microstates  (fuzzballs) have no horizon or singularity, and they do {\it not} collapse.  We consider two simple examples from classical gravity to illustrate how this violation of our intuition  happens.  In each case the `matter' arises from an extra compact dimension, but the topology of this extra dimension is not trivial. The pressure and density of this matter diverge at various points, but this is only an artifact of dimensional reduction; thus we bypass results like Buchadahl's theorem. Such microstates  give the entropy of black holes, so these topologically nontrivial constructions dominate the state space of quantum gravity.

\end{abstract}
\vskip 1.0 true in

\newpage
\setcounter{page}{1}

Suppose we collect together a sufficiently large mass $M$ of matter in a given region. According to classical general relativity,  the self-gravitation of this matter will pull it inwards,  resulting in a black hole.   Such a hole leads to Hawking's information paradox \cite{hawking}. In string theory, on the other hand, we find microstates for the black hole which have no horizon or singularity, and thus no information problem \cite{fuzzballs}.  What prevents these microstate configurations from suffering gravitational collapse?

The issue becomes sharper because of Buchdahl's theorem \cite{buchdahl}. If the matter is  a spherical ball of perfect fluid, then it cannot hold up against gravitational collapse once it is confined  within  a radius
\be
R_{min}={9\over 4}GM
\ee
If the ball is smaller than this, then  a straightforward computation  shows that the pressure $p$ diverges before we get to the center. 

The microstates of string theory -- often called fuzzballs -- are not made of perfect fluids, nor are they exactly spherically symmetric. Thus one cannot argue that they must be subject to Buchdahl's theorem.  But we must still understand how they hold up against the {\it spirit} of this theorem.  Intuitively, we might feel that  once an object   has a radius close to $r=2M$,  the gravitational pull will be  so strong that it must  collapse through a horizon. Is this intuition correct?

We will consider two examples where we have a shell with positive energy density, but where  this shell does {\it not} collapse inwards. These examples share some qualitative features with the construction of microstates in string theory, so they are a useful guide to the nature of fuzzballs. In each example we have an {\it extra dimension} compactified to a circle; dimensionally reducing on this circle gives Einstein gravity in 3+1 dimensions and a scalar field. The scalar field has positive energy density and gives the `matter' in the 3+1 dimensional description. 

\b

(a) For our first example, take flat 4+1 dimensional spacetime, where one direction has been compactified to a circle of radius $R$. Witten \cite{witten} showed that this spacetime was unstable to tunneling into a `bubble of nothing'. We are interested in the geometry after this bubble has nucleated:
\be
ds^2=-r^2 dt^2 + {dr^2\over 1-{R^2\over r^2}} + \cosh^2 t d\Omega_2^2 + (1-{R^2\over r^2}) d\psi^2
\label{metricb}
\ee
Here $\psi$ is the compact direction $0\le \psi\le 2\pi R$. Its proper radius goes to $R$ at infinity, and vanishes at $r\r R$. The entire spacetime is smooth however, as the $r, \psi$ directions form a `cigar', with a smooth tip to the cigar at  $r=R$. The region $r<R$ is {\it missing}; thus the angular sphere $\Omega_2$ at $r=R$ describes a `bubble of nothing'. If we dimensionally reduce on $\psi$, we get a scalar field. As we will see below in an analogous example, the energy density of this scalar field   increases to infinity as $r\r R$. But instead of collapsing inwards, we see from (\ref{metricb}) that the radius of the  bubble {\it expands} as $\cosh t$. 

\b

(b) As a second example, consider the 3+1 dimensional Euclidean Schwarzschild spacetime, and add in a time direction $t$
\be
ds^2=-dt^2 + (1-{r_0\over r})d\tau^2 + {dr^2\over 1-{r_0\over r}} + r^2 (d\theta^2+\sin^2\theta d\phi^2)
\label{metrickk}
\ee
Here the `Euclidean time' direction $\tau$ is compact, with $0\le \tau < 4\pi r_0$. The spacetime ends at $r=r_0$. The overall spacetime is smooth, as the $r_0, \tau$ directions form a cigar, whose tip lies at $r=r_0$.  If we dimensionally reduce on $\tau$, we get a scalar whose energy density diverges as $r\r r_0$. The spacetime is unstable to perturbations, but what is important for us is that it is time-independent; i.e., in spite of the energy density of the scalar $\tau$ in the 3+1 description, we do not see a gravitational collapse. What holds up the matter density?

\b

The physics is very similar in both cases, so we just  focus on (b). We write the 4+1 dimensional metric (\ref{metrickk}) as a scalar and a 3+1 dimensional metric. To do this we define
\be
g_{\tau\tau}=e^{{2\over \sqrt{3}}\Phi}
\ee
so that the scalar is
\be
\Phi={\sqrt{3}\over 2}\ln  (1-{r_0\over r})
\ee
The 3+1 dimensional Einstein metric is then defined as
\be
g^E_{\mu\nu}=e^{{1\over \sqrt{3}}\Phi} g_{\mu\nu}
\ee
which yields
\be
ds^2_E=-(1-{r_0\over r})^\h dt^2 + {dr^2\over (1-{r_0\over r})^\h} + r^2 (1-{r_0\over r})^\h(d\theta^2+\sin^2\theta d\phi^2)
\label{metric3}
\ee
With these definitions, the 5-d Einstein action gives
\be
S={1\over 16\pi G} \int  d^4 x   \sqrt{-g} \left ( R_E -\h \Phi_{,\mu}\Phi^{, \mu}\right )
\ee
so we get 3+1 Einstein gravity with a minimally coupled scalar. 
The stress tensor of the scalar field is then
 \be
 T_{\mu\nu}=\Phi_{,\mu}\Phi_{,\nu} -\h g^E_{\mu\nu} \Phi_{,\lambda} \Phi^{,\lambda}
 \ee
which works out to give
\be
T^\mu{}_{\nu} =  {\rm diag }\{-\rho, p_r, p_\theta, p_\phi    \}={\rm diag }\{-f, f, -f, -f    \} 
\ee
where
\be
f= {3r_0^2\over 8r^4 (1-{r_0\over r})^{3\over 2}}
\ee

Now we can investigate how the  the magic works, by looking at the metric (\ref{metrickk}) from the   3+1 dimensional perspective:

\b

(i) The energy density $\rho=f$ is positive. The radial pressure $p_r$ is positive as well, but the pressure in the angular directions $p_\theta, p_\phi$ is {\it negative}. Thus we do not have an isotropic pressure, while Buchdahl's theorem assumed an isotropic pressure. 

\b

(ii) The energy density  diverges as $r\r r_0$. The pressures diverge here as well. But these divergences are just an artifact of dimensional reduction: the full 4+1 dimensional metric is perfectly smooth. In Buchdahl's theorem, on the other hand,  a diverging density or pressure was regarded as a pathology.

\b

(iii) We find that $g_{tt}\r 0$ at $r\r r_0$, but $g_{tt}$ never changes sign. Thus there is no horizon. A geodesic starting at $r=r_0$ can escape to infinity in finite proper time.

\b

(iv) In the full 4+1 dimensional metric, the angular sphere had a nonzero proper radius $r_0$ at $r=r_0$. In the 3+1 dimensional metric (\ref{metric3}), the radius of the angular sphere goes to {\it zero} at $r=r_0$. Thus the surface $r=r_0$ in 4+ 1 dimensions looks like a {\it point} from the 3+1 dimensional perspective. Together with (iii), we can say that in the 3+1 dimensional language the point $r=r_0$ is one where a `horizon meets the singularity', so that we get a `naked singularity' rather than a black hole. But again, we should note that from the full 4+1 dimensional perspective, there is no horizon and no singularity. 

\b

Very similar features hold for the metric (\ref{metricb}).

\b

To summarize, the 3+1 dimensional description exhibits several pathologies like  divergent energy densities and pressures, while the full higher dimensional solution always remains regular. With such solutions, the intuition that positive energy density `pulls matter in' turns out to be incorrect, and we get static solutions like (\ref{metrickk}) or expanding solutions like (\ref{metricb}).

The significance of these observations is enormous. If we just look at 3+1 dimensional gravity plus normal matter, then it seems that black holes will necessarily form when enough matter collects together. The no-hair theorems tell us that we will then get a unique metric, and the information problem follows from Hawking's pair creation in this metric. But in string theory we find that there are {\it alternative} solutions with the same quantum numbers as the black hole. All these solutions have a nontrivial topology of the extra dimensions, similar to the toy examples we studied above. In \cite{gibbonswarner} it was shown how these higher dimensional solutions evade the no-hair theorems of the 3+1 dimensional theory. These solutions do not collapse under their own gravity, and they have no horizons or singularities as solutions of the full 10-dimensional theory. The black hole  is replaced by a linear combination of such `fuzzball' solutions, and we resolve the information paradox. 

 The early Universe has a singularity similar to the singularity of the black hole, and it is plausible that similar constructions will resolve the initial singularity \cite{mathurmasoumi}. Thus in situations where the energy is high enough to access these alternative topologies, our intuition about gravity derived from stellar models in inadequate, and we need a new perspective on the behavior of spacetime and matter.  

\section*{Acknowledgements}

This work is supported in part by DOE grant de-sc0011726, and by a grant from the FQXi foundation.

\end{document}